\documentclass[pdftex, 5p, preprint]{elsarticle}
\usepackage{amsmath}
\usepackage{amsfonts}
\usepackage{graphicx}

\newcommand{\h}{\Delta}

\newcommand{\erfc}{\mathrm{erfc}}

\newcommand{\uhp}{\mathbb{H}}

\begin{document}

\begin{frontmatter}
\title{Percolation and Schramm-Loewner evolution in the 2D random-field Ising model}
\author{Jacob D. Stevenson}
\ead{stevenso@uni-mainz.de}
\author{Martin Weigel}
\ead{weigel@uni-mainz.de}
\address{Institut f\"ur Physik, KOMET 331, Johannes Gutenberg-Universit\"at Mainz,
  Staudinger Weg 7, 55128 Mainz, Germany}

\begin{abstract}

  The presence of random fields is well known to destroy ferromagnetic order in Ising
  systems in two dimensions. When the system is placed in a sufficiently strong
  external field, however, the size of clusters of like spins diverges. There is
  evidence that this percolation transition is in the universality class of standard
  site percolation.  It has been claimed that, for small disorder, a similar
  percolation phenomenon also occurs in zero external field. Using exact algorithms,
  we study ground states of large samples and find little evidence for a transition
  at zero external field. Nevertheless, for sufficiently small random field
  strengths, there is an extended region of the phase diagram, where finite samples are indistinguishable from a critical percolating system. In this regime we
  examine ground-state domain walls, finding strong evidence that they are
  conformally invariant and satisfy Schramm-Loewner evolution ($SLE_{\kappa}$) with
  parameter $\kappa = 6$. These results add support to the hope that at least some
  aspects of systems with quenched disorder might be ultimately studied with the
  techniques of SLE and conformal field theory.

\end{abstract}

\begin{keyword}
SLE \sep percolation \sep random-field Ising model
\end{keyword}

\end{frontmatter}

%\maketitle
%\tableofcontents
%\bibliographystyle{naturemagurl}
%\bibliographystyle{aip}
%\bibliographystyle{/people/jake/research/apsrev}
%\bibliographystyle{apsrev}
\bibliographystyle{model1-num-names.bst}

The random field Ising model (RFIM) is one of the earliest studied and simplest
disordered systems showing non-trivial and glassy behavior
\cite{binder.1986,nattermann:97}. It has a number of important realizations in
nature, including diluted antiferromagnets in a field and binary liquids in porous
media \cite{nattermann:97}. Through its long history, researchers have managed to
gain a reasonable understanding of the critical behavior, although this progress has
been neither straight nor smooth, and many questions remain unanswered
\cite{nattermann:97}. It is known, for example, that the RFIM in two dimensions (2D)
lacks ferromagnetic order \cite{binder.1983,aizenmann:89a}. Even at zero temperature
it remains in the paramagnetic state for non-zero disorder. Numerical ground-state
calculations have shown, however, that even in the absence of a thermodynamic
transition there exists a geometric transition at which the size of the spin clusters
diverges in a manner bearing many similarities to classical site percolation
\cite{seppala.1998,seppala.2001}. While this transition is rather clearly established
in the presence of an external field, it has been argued that a similar percolation
phenomenon can also be observed in the absence of an external field for sufficiently
small disorder \cite{seppala.2001,kornyei.2007}. Here, we re-investigate the
zero-field behavior with large scale ground-state calculations, focusing on the
possible percolation phenomenon.

The observed relations to classical site percolation at this (non-zero or zero field)
geometrical transition motivate further questions of how far the similarities go.
Interfaces in two-dimensional percolation satisfy Schramm-Loewner evolution (SLE)
\cite{schramm:00,cardy:05}, but this property relies on conformal invariance which is
not conserved in the presence of disorder.  Recently, however, there have been
suggestions that the domain walls of certain other disordered systems satisfy SLE.
This intriguing possibility implies that conformal invariance, broken by disorder, is
restored, at least at criticality where relevant length scales diverge.

Schramm-Loewner evolution is a method for constructing a statistical ensemble of
curves in the plane from one-dimensional Brownian motion, thus classifying curves
with only one parameter, the diffusion constant $\kappa$ \cite{bauer.2006}.
Characteristic interfaces in many physical systems have been shown (in some cases
rigorously) to satisfy SLE$_\kappa$.  These include percolation ($\kappa=6$), self
avoiding walks ($\kappa = 4/3$), as well as spin cluster boundaries ($\kappa = 3$)
and Fortuin-Kasteleyn cluster boundaries ($\kappa=16/3$) in the Ising model. A number
of numerical studies have found interfaces in certain disordered systems consistent
with SLE, in particular the 2D Ising spin glass\cite{amoruso.2006,bernard.2007}, the
Potts model on dynamical triangulations \cite{weigel:05b}, the random bond Potts
model \cite{jacobsen.2009}, and the disordered solid-on-solid model
\cite{schwarz.2009}. Here we extend this list to include the 2D random field Ising
model.

In Sec.~\ref{sec:model} we introduce the RFIM and Schramm-Loewner evolution in more
detail, and discuss the approach used here for determining ground states of large
samples. Section \ref{sec:H0} is devoted to an investigation of the critical behavior
of the RFIM near the geometric transition, focusing on the behavior in zero external
field for small disorder. In Sec.~\ref{sec:direct_testing}, we report the results of
our tests of the correspondence between interfaces in the RFIM and Brownian motion
implied by SLE.

\section{ The model and method }
\label{sec:model}

We consider the random field Ising model in two dimensions with Hamiltonian
\begin{equation} 
  \mathcal{H} = - J \sum_{\langle i,j\rangle} s_i s_j - \sum_i h_i s_i.
  \label{eq:hamiltonian}
\end{equation} 
Here, the spins $s_i = \pm 1$ are located on the sites of a square lattice and
interact ferromagnetically with nearest neighbors. The local fields $h_i$ are
quenched random variables drawn from a normal distribution with mean $H$ and standard
deviation $\h$.  Since, at zero temperature, only the ratio $J/\h$ is relevant, we
take $J=1$ for simplicity. The spin-spin interaction $J$ induces a correlation
between the spins resulting in spin clusters which are compact up to a length
$\xi_b$. Above this scale the clusters are fractal objects, the magnetization is zero
and the system is paramagnetic.  As the randomness $\h$ is decreased, the breakup
length $\xi_b$ increases.  At and above three dimensions $\xi_b$ diverges at the
thermodynamic phase transition, below which the system is ferromagnetic.  In
two dimensions no thermodynamic phase transition exists, and $\xi_b$ diverges only at
$\h = 0$.  It has been argued that, for $H=0$, the breakup length scale $\xi_b$
increases with decreasing $\h$ as \cite{binder.1983}
\begin{equation}
  \xi_b \sim e^{A / \h^2}.
\end{equation}
Though there is no thermodynamic transition in 2D, the linear extent of the largest
clusters diverges for sufficiently large $H$. In most aspects, this divergence
appears to be consistent with standard site percolation
\cite{seppala.1998,seppala.2001,kornyei.2007}. It has been suggested that the
divergence occurs even at $H=0$ if $\h$ is below a critical value
\cite{seppala.1998,kornyei.2007}.  It is the characteristics of clusters of aligned
spins and their boundaries at this geometric transition which we focus on in
this study.

We restrict our investigation to ground-state spin configurations, which can be
efficiently constructed through a mapping to the well known minimum cut (or maximum
flow) problem in graph theory \cite{dauriac.1985,hartmann.2002}. Consider a directed
graph with $N+2$ vertices, and edges $(i,j)$ furnished with weights $c_{ij}$.  The
minimum $(s,t)$ cut is given by a subset of the edges of the graph, whose removal
disconnects vertices $s$ and $t$, such that the sum of the weights of the cut edges
is minimal.  Using the variables $x_i$, which are 1 if vertex $i$ is connected to $s$
and 0 otherwise, the total weight of a cut can be represented as
\begin{equation}
  C(\{\mathbf{x}\}) = \sum_{i,j=1}^N x_i (1-x_j) c_{ij}.
\end{equation}
With the identification of $x_i$ with spin variables (excepting the vertices $s$ and
$t$), and an appropriate choice of the weights $c_{ij}$, this function can be made to
precisely match the RFIM Hamiltonian (\ref{eq:hamiltonian}).  The minimum cut
separating the graph into vertices connected to $s$ and vertices connected to $t$
gives the minimum-energy way of cutting the RFIM lattice into clusters of up and down
spins, and thus corresponds to the ground state of equation \ref{eq:hamiltonian}.  The choice for the edge weights is, for $i \notin (s,t)$
\begin{equation}
  c_{ij} =  \left\{
    \begin{array}{ll} 0, &  i\ge j \\ 
      4J, & \mathrm{else}
    \end{array} \right. .
\end{equation}
For the edges connecting the ``spin'' vertices to $s$ and $t$, the
result is given in terms of the quantity $u_i = -h_i - \frac{1}{2} \sum_j
(c_{ij} - c_{ji})$ as
\begin{equation}
  c_{si} =  \left\{
    \begin{array}{ll} 0, &  u_i > 0 \\ 
      -u_i, & \mathrm{else}
    \end{array} \right.,\;\;\;
  c_{it} =  \left\{
    \begin{array}{ll} u_i, &  u_i > 0 \\ 
      0, & \mathrm{else}
    \end{array} \right. .
\end{equation}
$c_{is}$ and $c_{ti}$ are taken to be zero for all $i$.  Here, we use a fast
algorithm for solving the minimum cut problem based on the idea of ``augmenting
paths''\cite{kolmogorov.2003,kolmogorov.2004}.  The worst case scenario for the
running time of this algorithm is an unimpressive $O(N^3)$ (or more generally $V^2E$
where $V$ is the number of vertices and $E$ is the number of edges), however, the
algorithm was designed to optimize the typical case.  The optimization was carried
out for vision and image analysis problems, but even for the graph structure
corresponding to the RFIM ground state calculation the running time is proportional
to $N$ for the samples considered here.  In practice, the maximum system size is
limited more by computer memory constraints than by time.

Schramm-Loewner evolutions are defined in terms of a family $g_{t}$ of
conformal maps which take, formally, the upper half plane minus the curve
$\gamma_{t}$ (parametrized by ``time'' $t$) to the upper half plane, $g_{t}: \uhp
\setminus \gamma_{t} \to \uhp$.  This map can be defined (using complex notation and
suitable normalization) in terms of the differential equation
\begin{equation}
  \frac{\partial g_t(z)}{\partial t} = \frac{2}{g_t(z) - \xi_t}
\end{equation}
where $\xi_t$ is the unique driving function for the curve $\gamma_t$.  If the
curve (or the process generating the curve) satisfies SLE$_{\kappa}$, then
$\xi_t$ will be a Brownian motion with zero mean and variance $\kappa t$.
Numerically, rather than solving the differential equation, the map $g_t$ is
instead pictured as a series of maps $g_{i}$ which iteratively remove a small
section from the beginning of the curve.  To calculate the driving function
from a given curve, the incremental map $g_{i}$ is approximated using a
vertical slit map \cite{kennedy.2009}
\begin{equation}
  g_i(z) = i \sqrt{ -(z-\xi_i)^2 - 4 \Delta t_i } + \xi_i.
  \label{eqn:slit_map}
\end{equation}
The parameters $\xi_i$ and $\Delta t_i$ are determined from the coordinates of
the curve segment to be removed through the relations $\xi_i = x_{i,i-1}$ and
$\Delta t_i = y_{i,i-1}^2/4$.  More specifically, $x_{i,i-1}$ and $y_{i,i-1}$
are the coordinates of the $i$'th segment of the curve after undergoing the
$i-1$ successive maps $g_{i-1}\circ \ldots \circ g_{1}$.  
The parameter $\xi_i$ is the value of the driving function $\xi_t$ sampled at time
$t_i = \sum_{j\le i} \Delta t_j$.
The complex square
root in equation \ref{eqn:slit_map} is calculated, as usual, with the branch
cut along the negative real axis.

This is an iterative process in which the coordinates of the interface are
successively updated for each step, and thus the computational complexity is
$O(L_I^2)$ where $L_I$ is the length of the interface. If $L_I$ grows like
$\sim N^{d_f/2}$ (with $d_f = 7/4$ expected for percolation) in the number of
spins $N$, the resulting computational complexity is $O(N^{7/4})$, which is
significantly slower than the ground-state calculation which is $O(N)$ on
average.  We use a fast implementation of this ``zipper'' algorithm
\cite{kennedy.2007}, in which blocks of multiple slit maps are approximated
by a Laurent series.  Treating blocked maps in one step dramatically speeds
up the calculation such that it scales on average as $L_I^{1.3}$.  The loss in
accuracy from the approximation is minimal.

If the driving function $\xi_t$ can be shown to be Brownian motion then the
curves satisfy SLE$_\kappa$.  In practice, the finite size of the lattice and
the zipper algorithm introduces correlations between the increments of
$\xi_{i}$ and in their associated time values $t_i$. In particular the
distribution of time steps $t_i - t_{i-1}$ is highly non-trivial and has
significant correlations.  However, it should be emphasized that correlations
in the times $t_i$ at which $\xi_t$ is sampled does not imply correlations in
the underlying continuous driving function $\xi_t$.

\section{Phase diagram and behavior at zero field}
\label{sec:H0}

\begin{figure}[t]
\includegraphics[width=.48\textwidth]{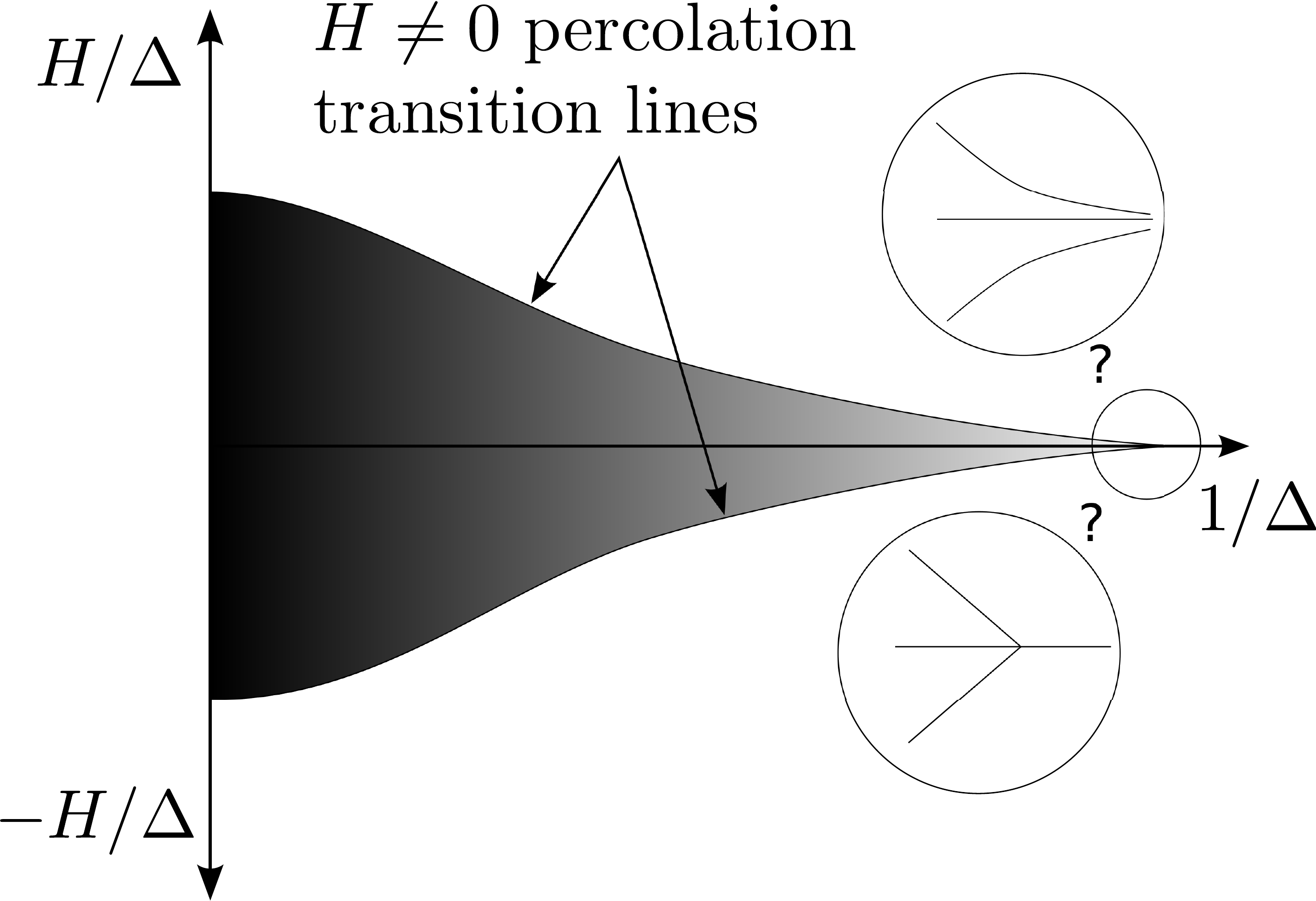}
\caption{Schematic representation of the dependence of the location $H_c$ of the percolation transition on
  the random-field strength $\h$.
  For $H<H_c$ the spin clusters have finite extent.  For $H\ge H_c$
  there exists at least one infinite cluster.
}
\label{fig:Hc}
\end{figure}

\begin{figure}[t]
\begin{tabular}{l}
\includegraphics[width=.48\textwidth]{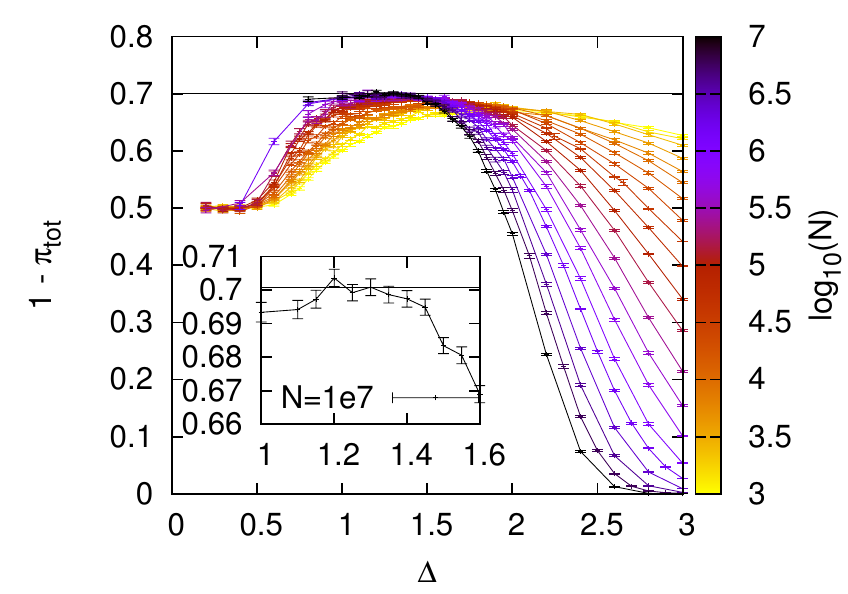} \\
\includegraphics[width=.48\textwidth]{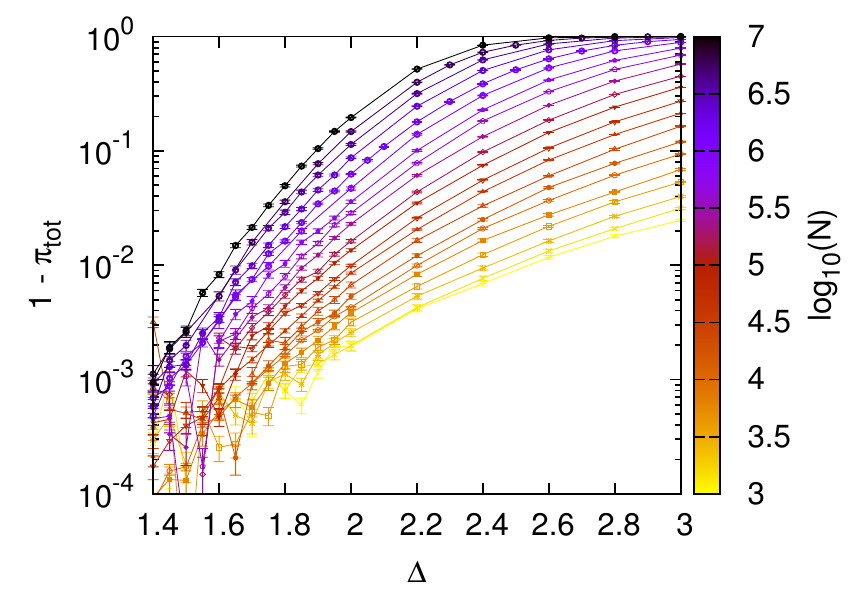}
\end{tabular}
\caption{ Crossing probabilities for spin clusters of the 2D RFIM on rectangular
  domains of aspect ratio $e^{-2/5}$ for a number of system sizes $N = e^{-2/5}L^2$
  at $H=0$. Top panel: probability of crossing of clusters of up spins in the
  vertical direction.  For $\h$ decreasing from infinity, the crossing probability
  approaches the exact percolation crossing value \cite{cardy.1992}, indicated by the
  horizontal line. At very small disorder, the breakup length $\xi_b$ is larger than
  the system size and the systems are effectively ferromagnetic. Inset: enlarged view
  of the plateau region for the largest system size.  The bottom panel shows
  $1-\pi_{tot}$, the total crossing probability.  If there was a transition at $H=0$,
  the lines would have to cross around the transition point.  }
\label{fig:h_dep}
\end{figure}

\begin{figure}[t]
  \includegraphics[width=.48\textwidth]{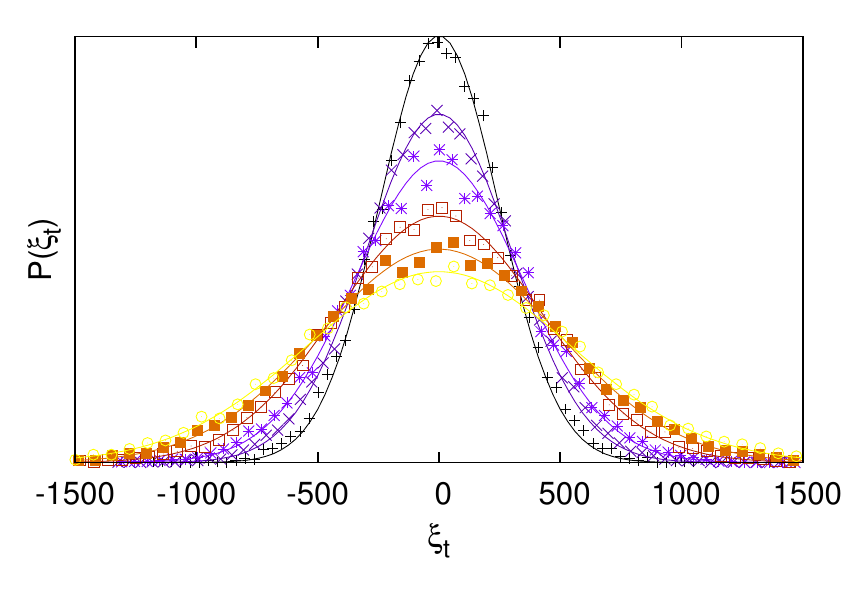}
\caption{
  Distribution of the random walk at several different ``times'' $t$ as extracted
  from the RFIM interfaces at $\h=1.65$.  The lines are normal distributions
  with zero mean and variance $\kappa t$ with $\kappa=6$.
}
\label{fig:SLEhist}
\end{figure}

For very large disorder, $\h \gg J z$, where $z$ is the lattice coordination number,
the interactions between the spins play no role and the system reduces, trivially, to
the classical site percolation model.  Each spin is determined solely by the
independent random variable $h_i$. Identifying spin up with ``site occupation'', the
site occupation probability $p$ is simply the probability for $h_i$ to be positive:
$p = \frac{1}{2} \erfc \{ -H / (\sqrt{2} \h) \}$.  For smaller disorder the spin-spin
interaction of the RFIM complicates the analogy with percolation. However, it has
been demonstrated that there exists a line of critical external fields $H_c(\h)$, as
pictured schematically in Fig.~\ref{fig:Hc}, for which observables like the crossing
probability and the fractal dimension of the spin clusters maintain the percolation
values \cite{seppala.2001}.

For very large disorder the line of critical external fields is found from the
critical site percolation probability $p_c$ ($p_c \approx 0.5927$ for the square
lattice) 
\begin{equation}
  H_c(\h \gg z) \approx -\h \sqrt{2} \erfc^{-1} \{2 p_c\}.
  \label{eqn:Hc_large_h}
\end{equation}
For small disorder the behavior is less well understood. $H_c(\h)$ decreases as
$\h$ decreases, approaching the limiting value $H_c(\h=0) = 0$.  It has been
claimed \cite{seppala.2001,kornyei.2007} that at a finite disorder strength,
$\h_c \approx 1.65 (5)$, the curve becomes identically zero $H_c(\h < \h_c)=0$,
cf.\ Fig.~\ref{fig:Hc}.  We tested this claim by looking at the behavior of the
system varying $\h$ at $H=0$.  We calculated the probability that a connected
cluster of up spins crosses a rectangle of aspect ratio $e^{-2/5}$ in the vertical
(shorter) direction (this aspect ratio is chosen to be different from $1/2$, but
the specific value is somewhat arbitrary). In the upper panel of
Fig.~\ref{fig:h_dep}, we present the
results of these calculations. The crossing probability $\pi$ for large $\h$ is
zero (as expected for a square lattice with $p_c > 0.5$).  As the disorder is
decreased, $\pi$ approaches the exact percolation value, which is indicated by
the horizontal line in the plot.  At very small disorder, when the breakup
length scale $\xi_b$ becomes comparable to the system size $L$, the system
appears ferromagnetic and the crossing probability falls away from the plateau
value.

To separate the approach to the plateau from the finite size effects we also
considered another quantity, the probability $\pi_{tot}$ that there exists a
connected cluster of either up or down spins crossing either horizontally or
vertically.  The small disorder limit of this quantity is $\pi_{tot}(\h \to 0) = 1$
for both the onset of percolation and the ultimate ferromagnetic (and finite size)
ordering.  If there is a percolation transition at finite disorder ($\h_c >0$), in
the thermodynamic limit $L=\infty$ the curve $\pi_{tot}(\h)$ will be a step
function. Hence, the curves for finite systems approaching this step function will
intersect at $\h_c$. As is apparent from our data for $\pi_{tot}$ shown in the lower
panel of Fig.~\ref{fig:h_dep}, such a crossing is not observed at least down to $\h
\approx 1.45$, significantly below the previously conjectured value of $\h_c \approx
1.65$. The study of even smaller disorder strengths, while ensuring $L \gg
\xi_{b}$, is preempted by the exponential
growth of the breakup length $\xi_b$. Although it seems very unlikely that the system undergoes a true
percolation phase transition for $H=0$, there exists a large plateau region in which
the system appears to be at critical percolation at $H=0$, even up to very large
system sizes. Presumably, this holds true even for system sizes occurring in
experimental realizations of the RFIM.

\section{Schramm-Loewner evolution}
\label{sec:direct_testing}

Finally, we tested directly the conformal mapping upon which SLE is based.  We
studied the statistics of $10\,000$ interfaces generated in a half disc.  This
geometry is used to optimally mimic the full half plane.  The interface is initiated
at the origin (the center of the flat edge) by two fixed spins and is considered
ended when it touches the curved boundary.  We find that the variance of the driving
function calculated from the interfaces using the method described in
Sec.~\ref{sec:model} is $\hat{\kappa} = \langle (\xi_t - \langle \xi_t \rangle )^2 \rangle/t = 6.09 \pm 0.09$, and
the normalized mean is $\hat{\xi} = \langle \xi_t \rangle / \sqrt{ \hat{\kappa} t } = 0.01 \pm
0.01$.  The agreement with SLE is good, but is only expected to be perfect at
criticality.  The difference from the expected percolation value $\kappa = 6$ which
is perhaps just visible here is due to the calculations being carried out at
$\h=1.65$.  The position distribution of the resulting stochastic process at several
fixed times is shown in Fig.~\ref{fig:SLEhist} along with curves representing the
expectations for a perfect random walk.

We have focused here on the case of zero external field, where we have shown,
using significantly larger system sizes than had been accessible before, that
there is no percolation phase transition for $H=0$, at least down to $\h =
1.45$. Due to the fact, however, that the percolation transition line comes
exponentially close to $H=0$ for small random-field strengths $\h$ (cf.\
Fig.~\ref{fig:Hc}), there exists a plateau region where even at $H=0$ the
behavior appears nearly indistinguishable from criticality.  In this regime we
have tested the applicability of Schramm-Loewner evolution to the RFIM and
found good agreement. Further studies at non-zero external field and using an
array of further tests for consistency with SLE confirm this result
\cite{stevenson:prep}.

The authors acknowledge computer time provided by NIC J\"ulich under grant No.\ hmz18
and funding by the DFG through the Emmy Noether Program under contract
No. WE4425/1-1.

%\bibliography{SLE_RFIM}

\end{document}